\documentclass[tighten, times, twocolumn]{aastex631}  
\shortauthors{Mould \& Thakore}
\usepackage{graphicx}
\usepackage[mathlines]{lineno}
\usepackage{color}
\usepackage{amsmath}
\usepackage{xcolor}
\newcommand{\kms}{\mbox{km\,s$^{-1}$}}
\newcommand{\etal}{\mbox{\rm{et al.}~~}}

\begin{document}

\title{Primordial Black Holes as Coma Cluster Dark Matter
and the Unresolved $\gamma$-Ray Background}

 \shorttitle{PBH as cluster DM}
\author{Jeremy Mould}
\affiliation{Swinburne University}
\affiliation{ARC Centre of excellence for Dark Matter Particle Physics}
\author{Bhashin Thakore}
\affiliation{Dipartimento di Fisica, Universit\`a degli Studi di Torino, via P. Giuria 1, 10125 Torino, Italy}
\affiliation{INFN -- Istituto Nazionale di Fisica Nucleare, Sezione di Torino, via P. Giuria 1, 10125 Torino, Italy}



\begin{abstract}
If 0.1\% of the dark matter in the Coma cluster is constituted by primordial black holes (PBHs) with masses ranging from 1 $\times$ 10 $^{-19}$ M$_\odot$   to 100 $\times$ 10$^{-19}$
 M$_\odot$, then the observed GeV $\gamma$-ray emission from the cluster could potentially be attributed to Hawking radiation. This process describes the thermal emission of particles by black holes due to quantum effects near their event horizon. The emitted spectrum is inversely proportional to the black hole's mass, meaning lighter PBHs radiate at higher energies, potentially falling within the GeV range. If 0.1\% of
 the Coma cluster's dark matter is PBH in this mass range, a fit to
 the cluster's GeV emission is obtained. We then investigate the potential for constraining evaporating primordial black holes (PBHs) through cross-correlations between the Unresolved Gamma-Ray Background (UGRB) and weak gravitational lensing. Utilizing 12 years of Fermi-LAT observations and weak lensing measurements from the Dark Energy Survey Year 3 (DES Y3), we assess whether such correlations can reveal a PBH-induced $\gamma$-ray component. While a statistically significant correlation between the UGRB and large-scale structure has been observed (reaching a signal-to-noise ratio of 8.9 in recent analyses), this signal is consistent with emission from clustered astrophysical sources such as blazars. We find that attributing a measurable fraction of the UGRB to PBH evaporation would require unrealistically large PBH abundances. Our results indicate that, given current sensitivity and modeling assumptions, this cross-correlation approach alone is insufficient to robustly constrain PBH evaporation. In a separate line of inquiry, we draw attention to a cluster of Coma-like X-ray clusters, designated Draco X, observed at a redshift of z = 0.12. These systems, characterized by their significant X-ray emission from hot, diffuse intracluster gas, represent massive gravitationally bound structures. The existence and properties of such clusters at these redshifts provide crucial cosmological probes, offering insights into the formation and evolution of large-scale structure and the underlying cosmological parameters. Further investigation of Draco~X and similar high-redshift clusters would yield additional constraints on large scale structure. 
\end{abstract}
 \section{Introduction}
 Clusters of galaxies were the first entities in which dark matter (DM) was recognized, and the Coma cluster was the exemplar
 site \citep{z}. 
 Despite the dominant role of dark matter in galaxy clusters, its composition remains unknown. While weakly interacting massive particles (WIMPs) have historically been a leading candidate, the lack of direct detection has led to increasing interest in alternative scenarios, including Primordial Black Holes (PBHs). If PBHs contribute significantly to cluster-scale dark matter, their interactions with the intracluster medium (ICM) via Hawking radiation could have observable consequences. The evaporation of PBHs through Hawking radiation produces a continuous spectrum of high-energy particles, including photons, electrons, and neutrinos (\cite{hawking1974black}). This process is particularly significant for PBHs with masses below $10^{15}$ g, which can evaporate within the age of the Universe. In dense environments such as galaxy clusters, the accumulation of PBH-induced emissions may contribute to the diffuse high-energy background, offering a potential observational signature distinct from other astrophysical sources. In addition to their possible contribution to high-energy emissions, PBHs in clusters may also play a role in shaping large-scale structure. If PBHs were formed in the early Universe from primordial density fluctuations, their spatial distribution could be correlated with that of massive halos. This clustering behaviour could enhance their detectability when tracers of large-scale structure such as weak lensing or galaxy clustering are cross-correlated with diffuse high-energy backgrounds such as the Unresolved Gamma-Ray Background (UGRB). 
 
 The fraction of that DM that is primordial black holes (PBH) is unknown. With the advance of microlensing capabilities, the portion that is subsolar mass may soon be measured. \cite{BM} 
 have considered the ionization effectiveness of Hawking radiation in the context of the Hubble tension. Here we examine what contribution Hawking radiation might make to the Coma cluster's high energy emission at GeV energies and above ($\S$2 and $\S$3). 
 At degree resolution ($\S$4) such clusters may contribute to the Unresolved $\gamma$-ray
 Background\footnote{At redshift 10 a fully developed Coma cluster would be 4$^\prime$ in diameter.}. Finally, we examine the large scale structure of clusters like Coma at higher redshifts ($\S$5.)




 \section{The physical environment }
The Coma cluster is a colossal cosmic structure, with its baryonic mass dominated not by stars, but by an extensive reservoir of X-ray emitting gas. This hot, diffuse plasma accounts for approximately 10$^ {14}$ M$_\odot$, a mass six times greater than that locked within its stellar populations. The primary constituent of this intracluster medium (ICM) is a tenuous plasma of baryons heated to extreme temperatures, on the order of $\sim$10$^8$K. At these temperatures, the gas emits profusely in the X-ray band via thermal bremsstrahlung. While the bulk of the X-ray emission arises from this hot ICM, \cite{ch} 
further highlight the potential presence of warm-hot intergalactic medium (WHIM) filaments. These filaments, predicted by cosmological simulations, are thought to connect galaxy clusters and host baryons at temperatures between 10$^5$
  and 10$^7$K, representing a significant component of the "missing" baryons in the universe.

Despite the substantial amount of hot gas, the Coma cluster's ICM remains optically thin to electron scattering. This is a crucial point for observations of high-energy phenomena, as it implies that the GeV emission originating from hypothetical 10 $^{-17}$ M$_\odot$ PBHs would not be significantly attenuated or distorted by scattering off free electrons within the cluster. The optical depth to electron scattering, often denoted as $\tau$, is given by $\tau~=~\int \sigma n_e dl$, where n$_e$
is the electron number density, $\sigma$ 
  is the Thomson scattering cross-section, and the integral is along the line of sight. For Coma, this value is sufficiently low that it preserves the direct path of any emitted GeV photons.

The Planck collaboration \cite{ade2013planck}  
has meticulously mapped the Compton y-parameter for the Coma cluster. The y-parameter is a dimensionless quantity that quantifies the distortion of the cosmic microwave background (CMB) spectrum due to the Sunyaev-Zel'dovich (SZ) effect. This effect arises from the inverse Compton scattering of CMB photons off the hot electrons in the ICM, resulting in a measurable shift in the CMB photon energies. Crucially, the y-parameter, along with the optical depth to electron scattering and the dispersion measure (relevant for radio observations, representing the integral of the electron density along the line of sight,  are all direct line integrals. Specifically, the y-parameter is proportional to $\int n_e T  dl$, where T is the electron temperature, highlighting its sensitivity to both the density and temperature of the hot gas. These integrated quantities provide valuable constraints on the physical properties and distribution of the ICM. 
  
\begin{figure*}
	\includegraphics[width=0.5\textwidth]{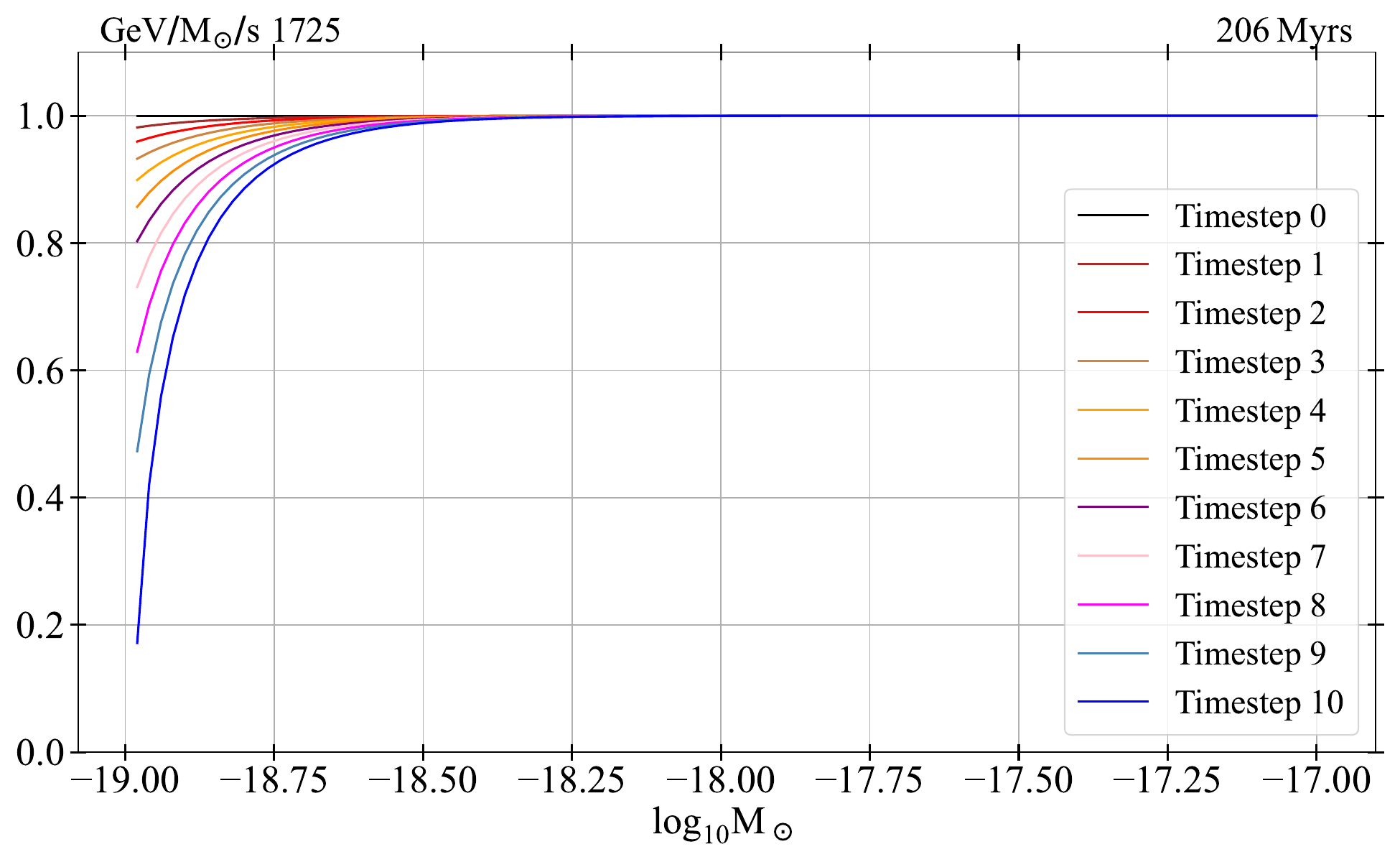}
	\includegraphics[width=0.5\textwidth]{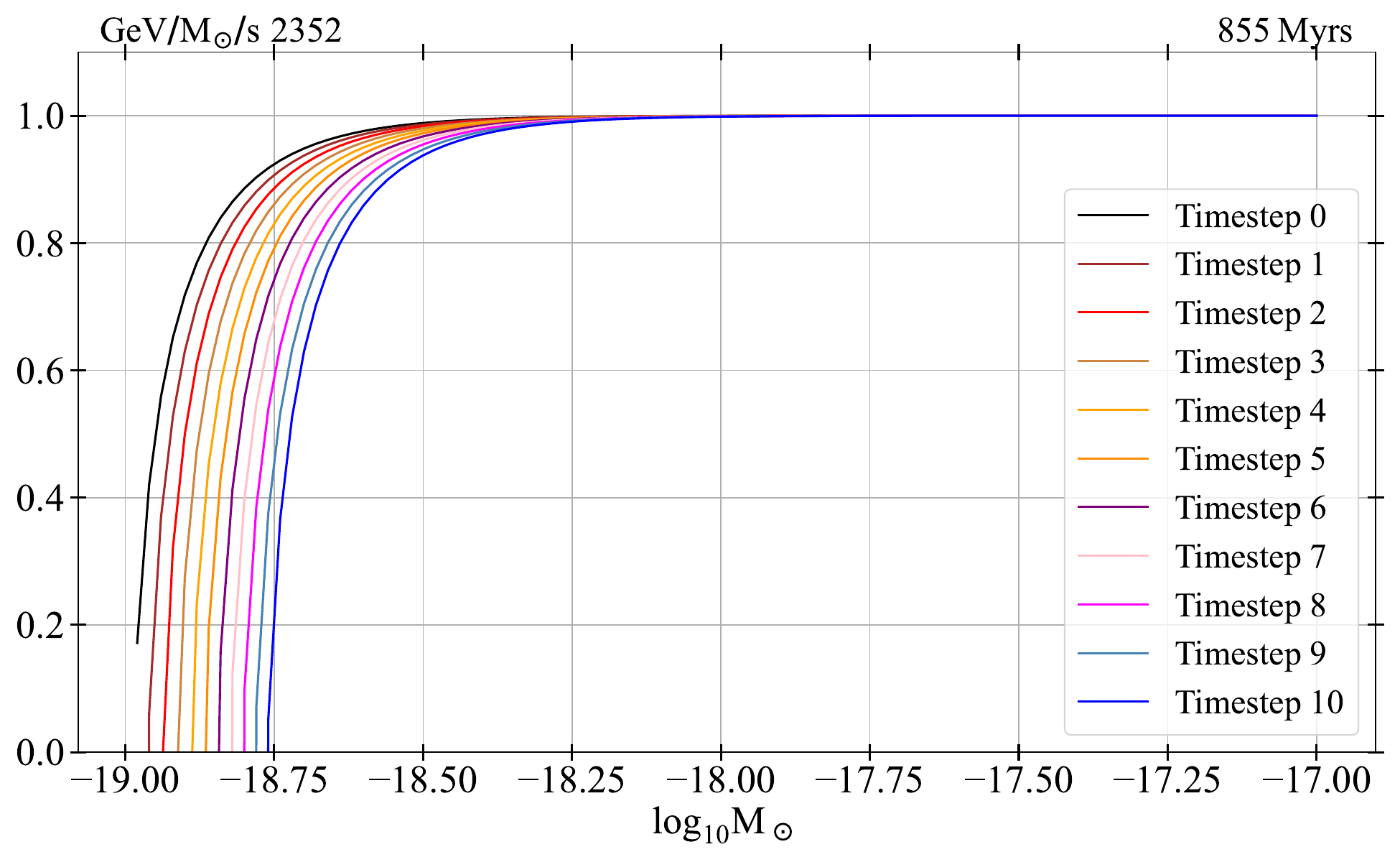}
    \\
	\includegraphics[width=0.5\textwidth]{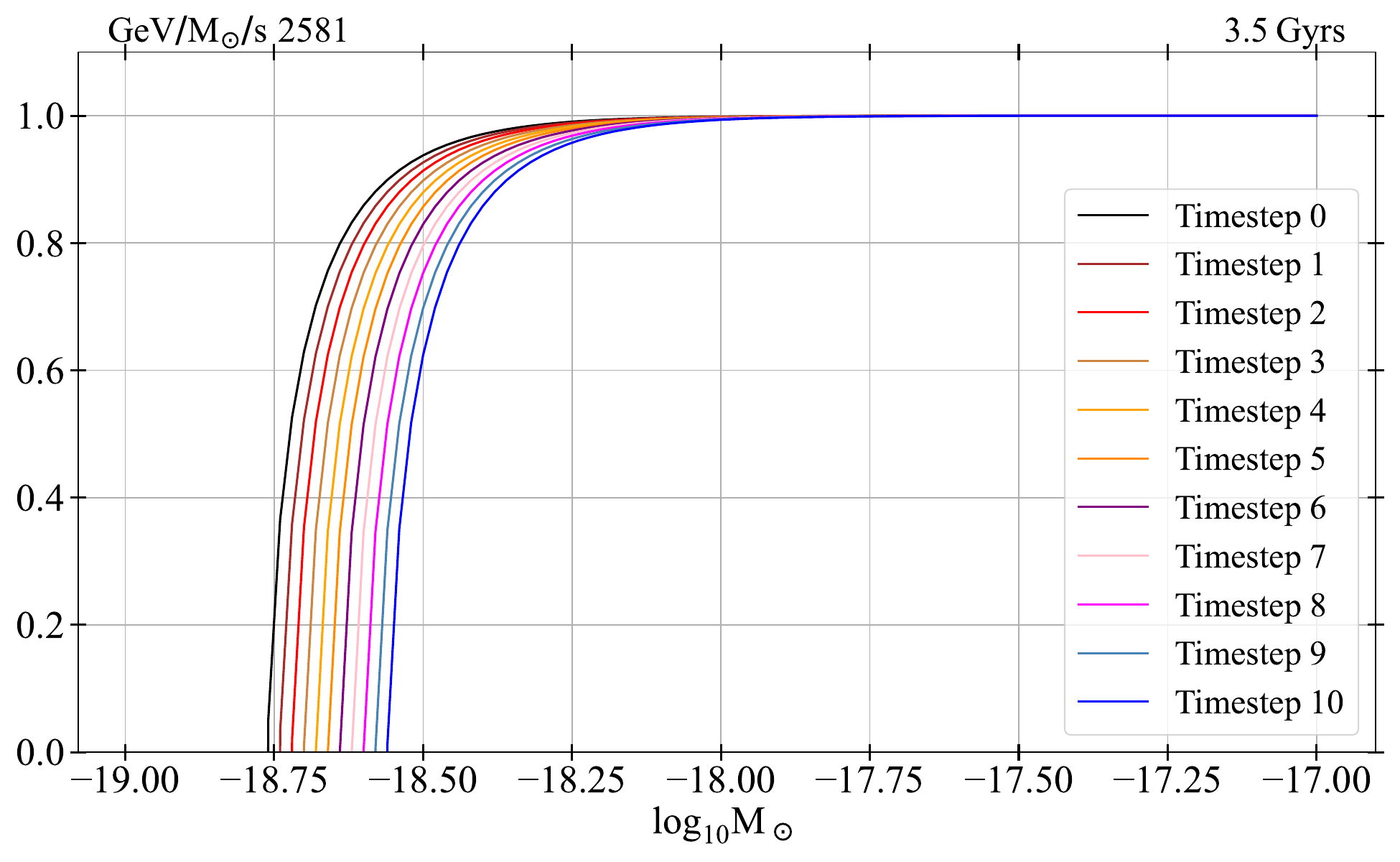}
	\includegraphics[width=0.5\textwidth]{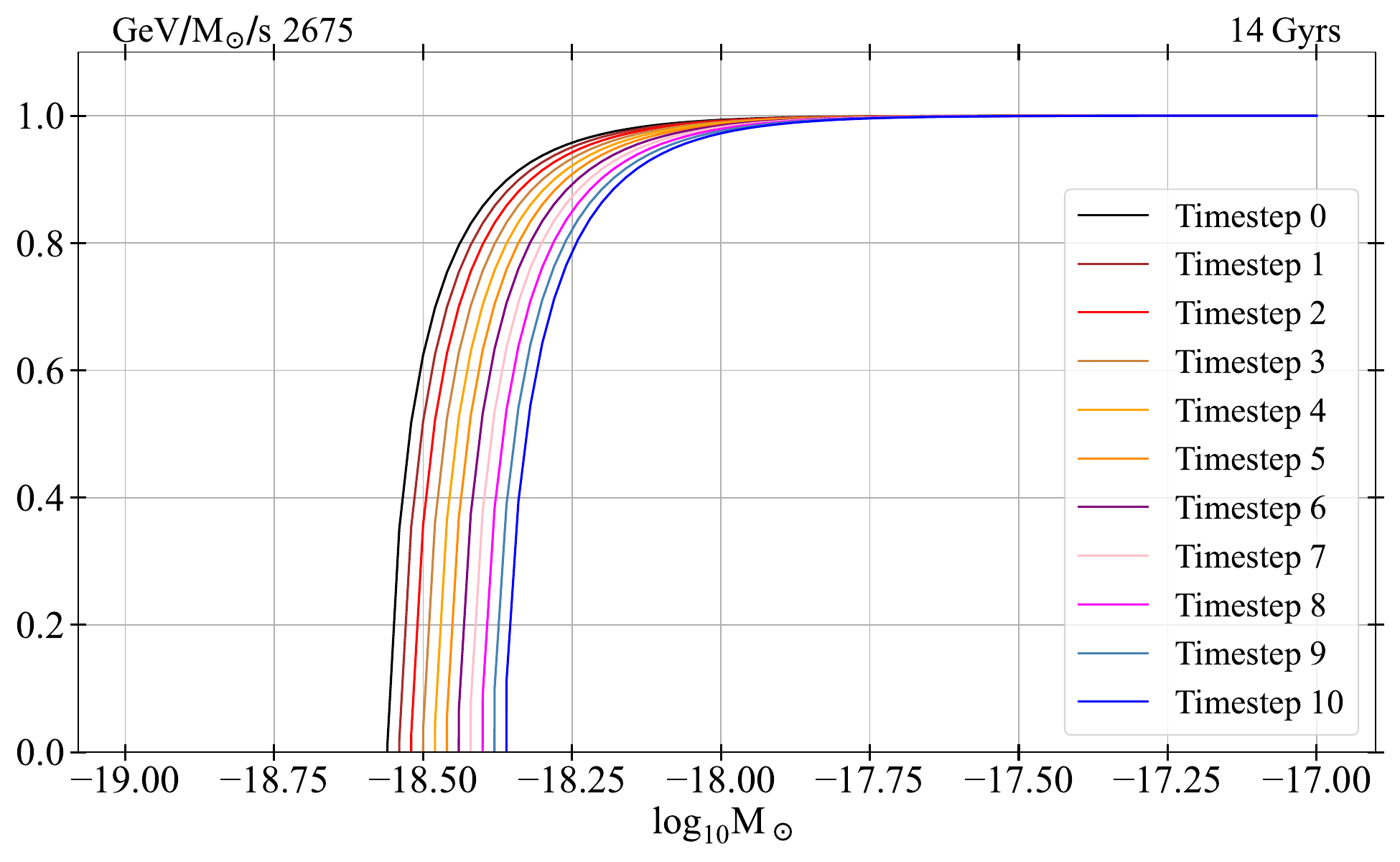}
	\caption{ The proportion of PBH fuel consumed as a function of black hole mass after 206 Myr (top left), 855 Myr (top right), 3.5 Gyr (bottom left) and 14 Gyr (bottom right). In each subplot, the x-axis is the PBH mass bins and the y-axis is the fraction of total PBH mass remaining after 10 timesteps. The coloured lines indicate the sequence of timesteps with the first in black and the last in blue, as seen in the vertical colour bar in the first subplot. 
    As time passes,
	progressively more of the initial PBH reservoir is drained, with age
	at the end of the numbered subplot in Myrs and Gyrs and the power in GeV/s
	as Hawking radiation yield per solar mass of DM. If the energy of
	the cutoff were detected at a local redshift, it would give an age
	dependent on h, k, and G, rather than on the expansion of the Universe.}
    \label{fig:PBH_evaporation}
\end{figure*}

 Using the PBH mass loss rate of \cite{mos}, 
  and dividing the mass range 10$^{-19}$ to 10$^{-17}$ M$_\odot$ into 100 increments in log mass, we show the evaporation of PBH as a function of mass and time in Figure \ref{fig:PBH_evaporation}.

The mass loss includes both Compton scattering and pair production, which energize electrons and thus heat the gas, helping to maintain its 10$^8$K temperature.
For every gram of baryons, there is
\[
\frac{0.12}{0.0224} \times f \, \text{gm of PBHs},
\]
which yields an energy injection rate of
\[
\frac{5.4 f}{m} \frac{dm}{dt} \, \tau c^2 \quad \text{ergs/s},
\]
where \( \tau \) is the optical depth in the relevant particle channels.

The fractional mass loss rate is
\[
\frac{1}{m} \frac{dm}{dt} = \alpha \frac{\hbar c^4}{G^2 m^3} = 3 \times 10^{-23} \, \text{s}^{-1}.
\]
So the yield becomes
\[
0.15 \, \tau \times 27 \times 10^{-3} \quad \text{ergs/gm/s}.
\]

The heat capacity of a monatomic gas is
\[
\frac{3k}{2} = \frac{1.38 \times 10^{-16}}{\mu \cdot 1.6 \times 10^{-24}} = 10^8 \quad \text{ergs/gm/K},
\]
where \( \mu \) is the mean molecular weight in units of the proton mass.

Thus, the temperature increase from PBH pair production is negligible.
Keeping the aforementioned physical environment in mind, we can model the sources of the emission in the next section.
 \section{Source of the emission \label{sec:srcemission}}
\begin{figure*}
	\includegraphics[width=0.5\textwidth]{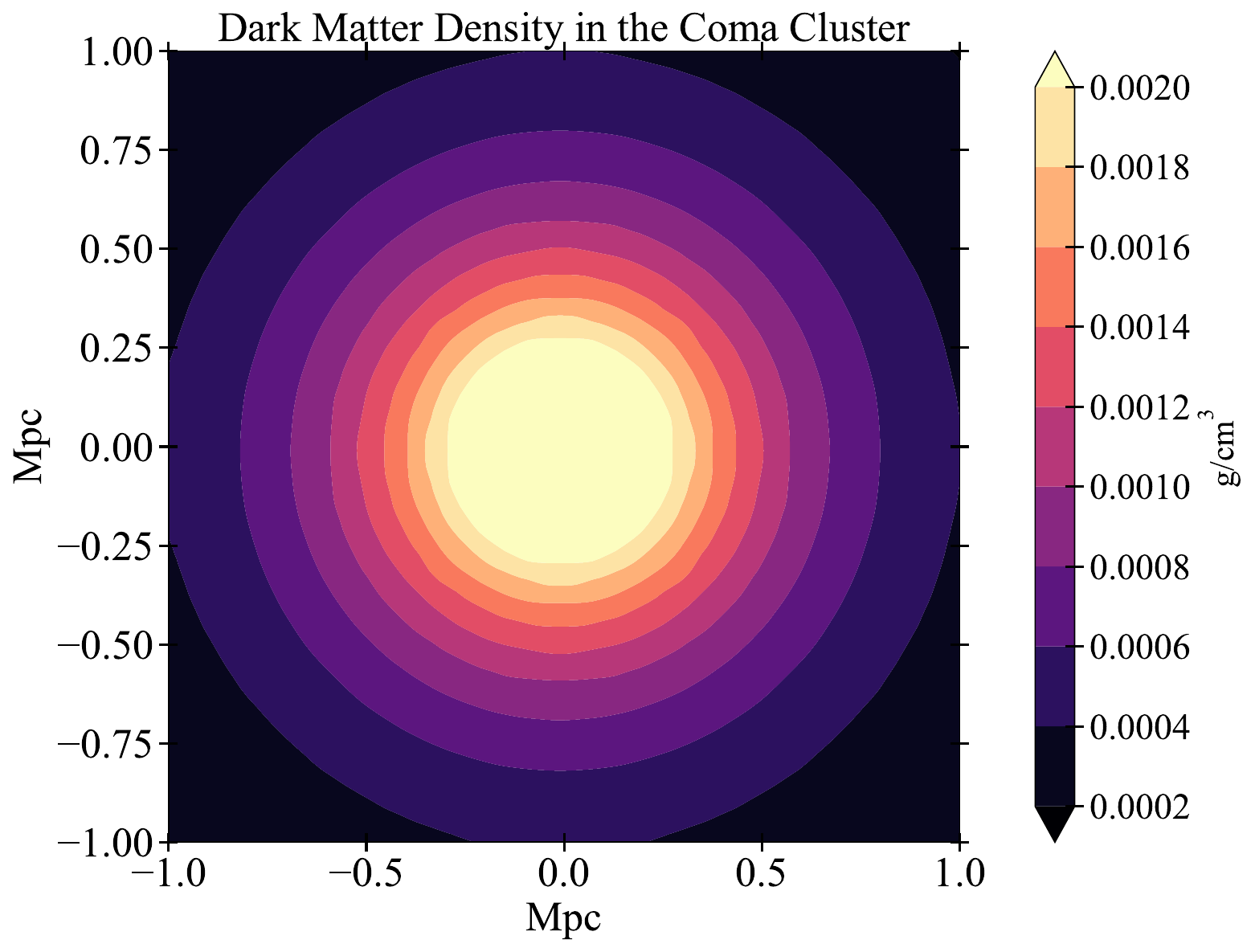}
	\includegraphics[width=0.5\textwidth]{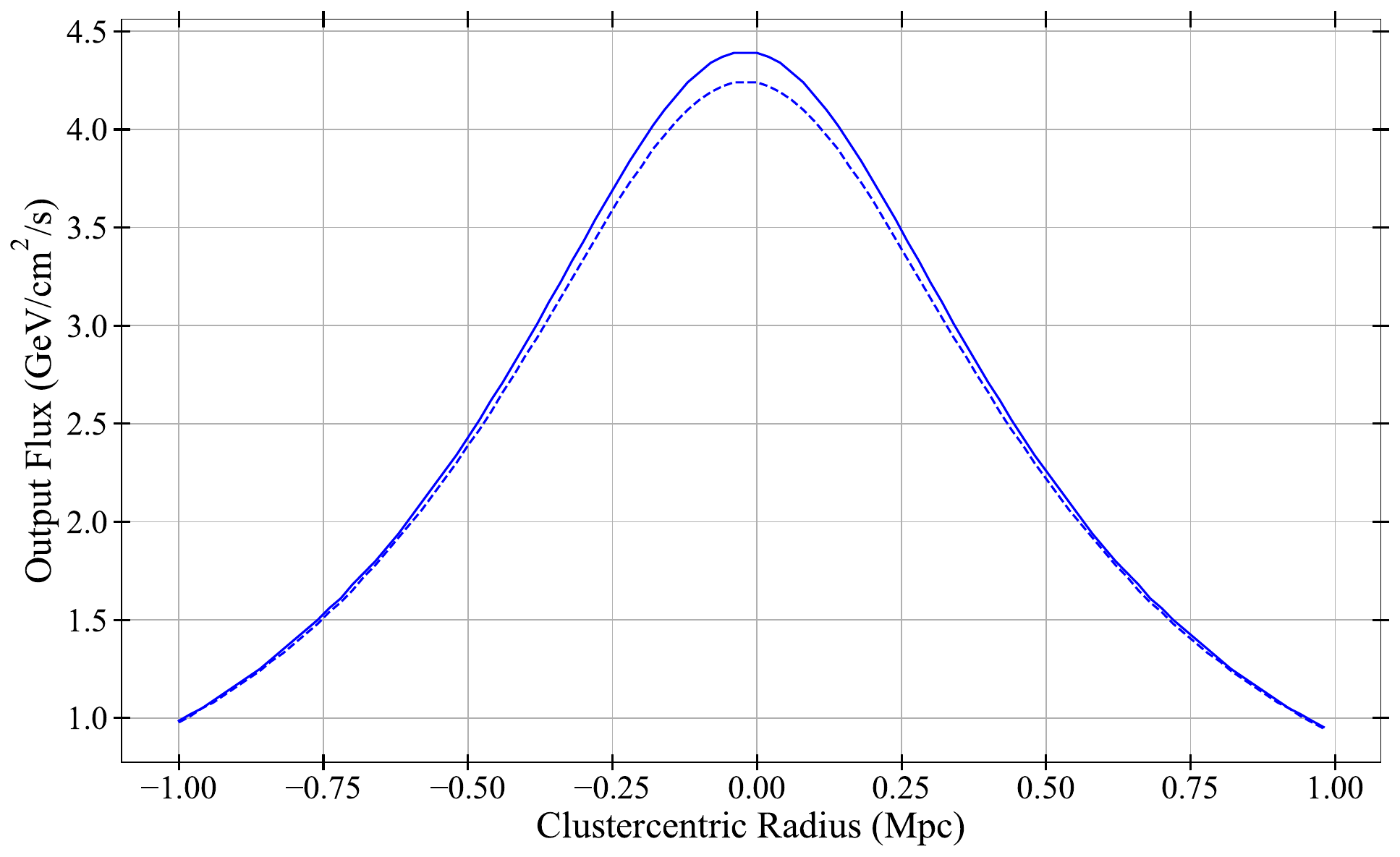}
    \\
	\includegraphics[width=0.5\textwidth]{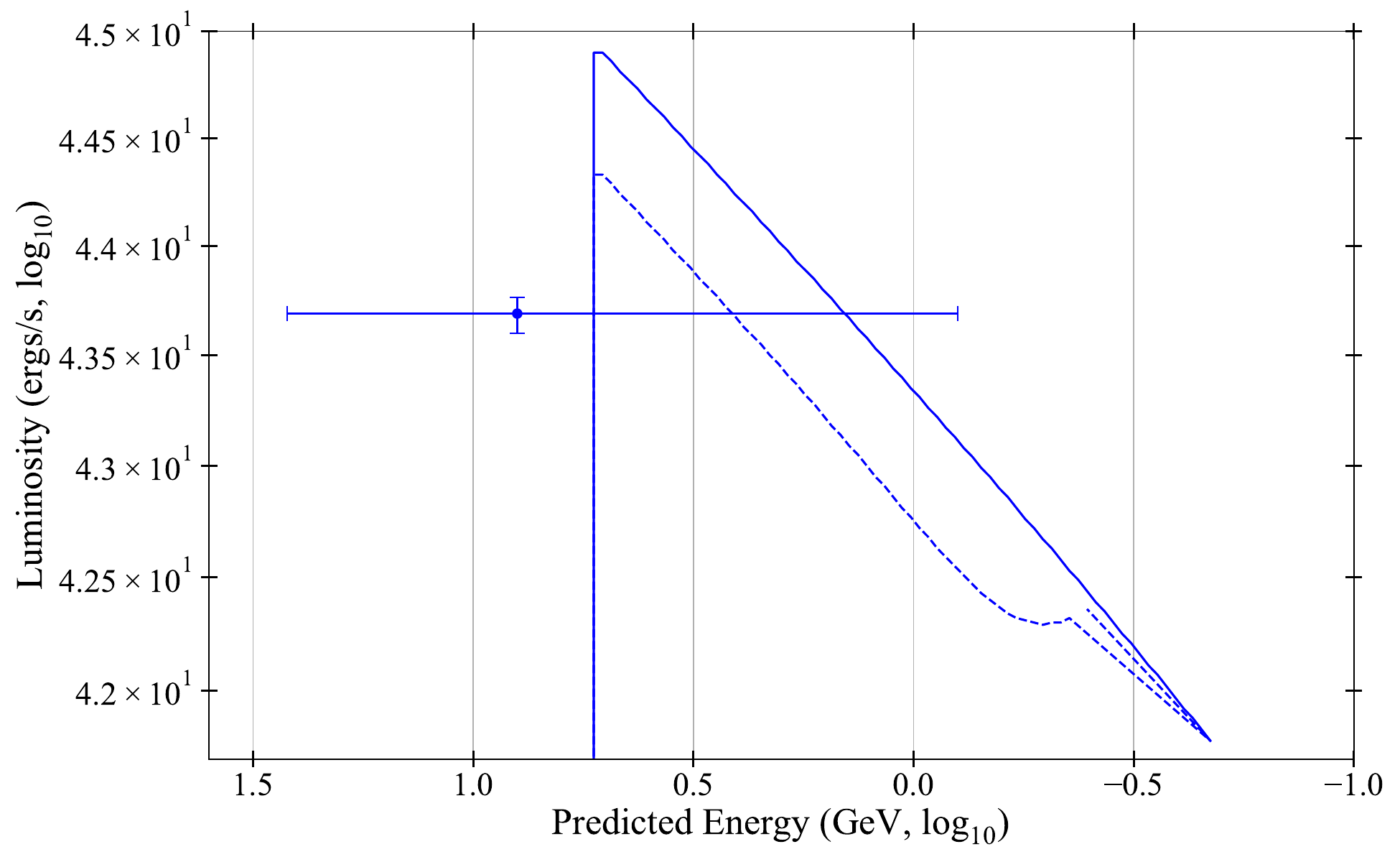}
	\includegraphics[width=0.5\textwidth]{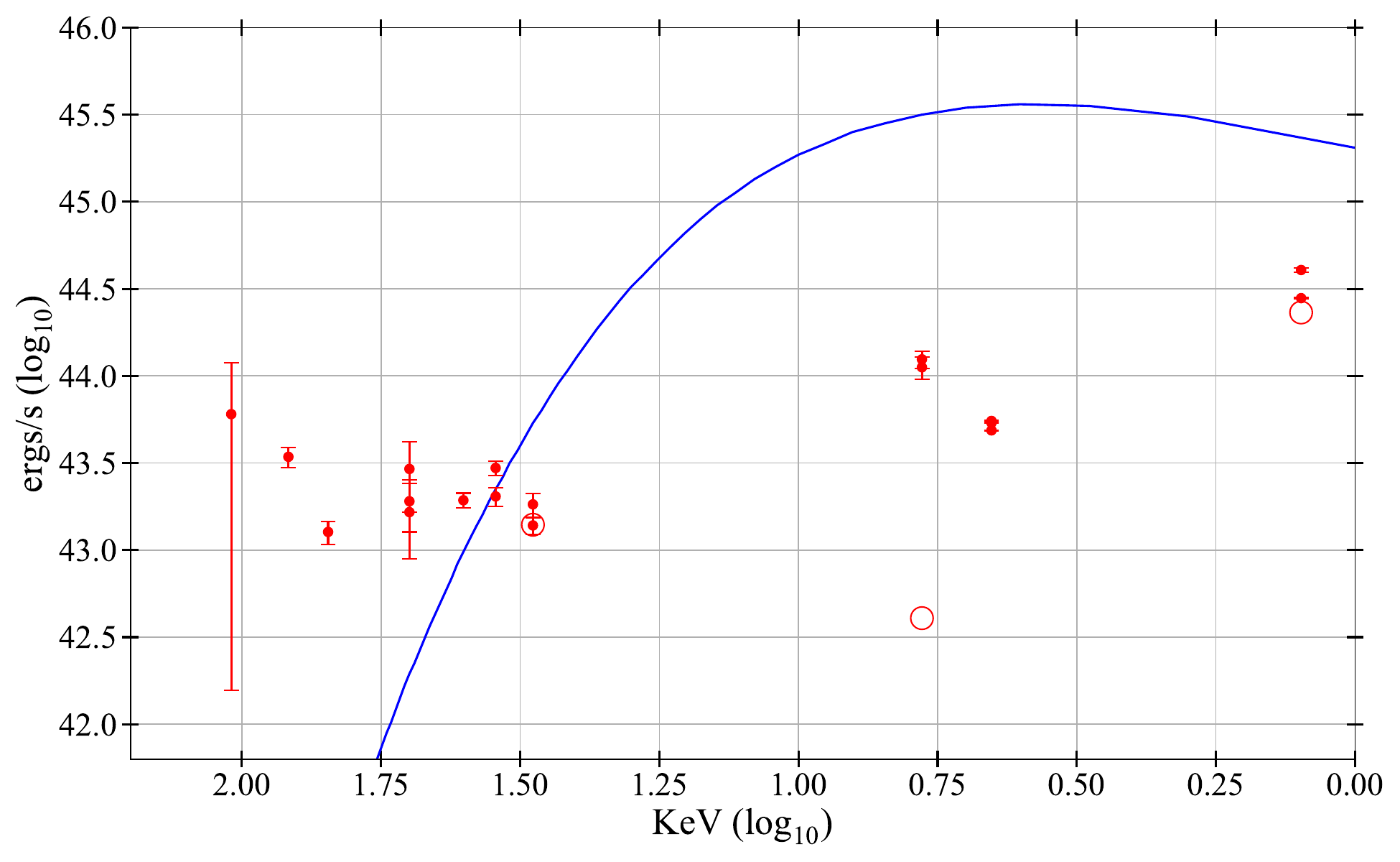}
	\caption{Coma's high energy spectral energy distribution. Top left:
	contours of dark matter density. The outer contour
	is 0.0002 gm/cm$^3$ and the inner area $>$ 10 times that. 
	Top right: the output flux as a function of the clustercentric radius.
	The dashed curve shows the loss due to pair production at 5 GeV.
	Bottom right: Uncertainties are taken from the reference in Table 1. Open circles are measurements that seem discordant. The solid line is from Planck's Model D. Bottom left: the predicted GeV emission
	with the observation of \cite{baghmanyan2022detailed} shown as an error bar
	with extended energy coverage. The downward slope of the prediction
	with decreasing energy reflects the assumed PBH IMF. The cutoff
	just above 5 GeV is due to exhaustion of lower mass PBH, whose mass loss rate
dm/dt 	is proportional to m$^{-2}$.}
\label{fig:coma_spectral}
\end{figure*}

 With the assumption that the DM and hot baryons have the same distribution, we
 show  contours of GeV flux 
 in Figure \ref{fig:coma_spectral} (top left). This assumption is an approximation. The main violation of the assumption is the ability of star formation and especially supernova explosions to drive outflows of mostly baryons into the ICM.
An upper limit is  by \cite{fb} 
: 0.015 M$_\odot$/yr/10$^9$ L$_\odot$, which is 0.015 $\times$  10$^{14}$ M$_\odot$/Gyr $<<$ 10$^{16}$    M$_\odot$ of cluster mass.

Coma is optically thin to both electron scattering and pair production. 
The initial mass function (IMF) for PBH was
taken to be n $\sim$ m$^{-1}$ \citep{jrm} 
, and $\Omega_c/\Omega_b$ = 0.120 / 0.0224
for the DM to baryon ratio from the Planck collaboration (2020).
 To fit the available GeV data, 
 we set the fraction of DM that lies between 
 10$^{-19}$ and 10$^{-17}$ M$_\odot$ to f = 10$^{-3}$.
\cite{kush} 
 propose secondary emission from a giant radio halo
 as the source of the GeV emission measured by \cite{baghmanyan2022detailed}
 from 200 MeV to 300 GeV. The DM fraction in our Hawking radiation model 
 is therefore an upper limit, f $<$ 10$^{-3}$. The fraction that may be synchrotron
 radiation depends on the magnetic field present, B$^2$, and the uncertainty in
 the expected flux from 2$\delta$B/B is 68\% \citep{bon} .

 Figure \ref{fig:coma_spectral} also shows Planck Model D for n$_e$, T, and
 the free-free emissivity of \cite{pon} 
 with Gaunt factor unity. A distance of 100 Mpc was adopted for the Coma cluster (\cite{scolnic2025hubble}).
The available x-ray data is shown\footnote{Compiled by NED [ned.ipac.caltech.edu],  which also catalogs the sources (e.g. \cite{bird}, 
\cite{ff}, 
\cite{oh} 
and mission names (e.g. Swift, ROSAT, Beppo-SAX, INTEGRAL). } in Table 1. 
The free-free emission model in the keV regime is a simple model, but it is not a
very good fit to the data, which seems to be a flatter spectrum. In the GeV regime of the potential Hawking radiation, there is little data, but what there is, is in satisfactory agreement with the Hawking radiation hypothesis.
\begin{table}[h]
\caption{\bf Coma cluster x-ray flux}
\begin{tabular}{llrl}
\hline
\#&    Bandpass &     Flux~~~~&Reference\\
\hline
       1$^1$& 14-195  Swift  &  4.75E-11 &2018 ApJS, 235, 4\\
       2$^2$& 14-195  Swift &   3.59E-11 &2010 ApJS, 186, 378\\
       3$^3$& 15-150  Swift &   2.70E-11 &2010 A\&A, 524, 64\\
       4$^2$& 40-100  INTEGRAL&   1.00E-11&2006 ApJ, 638, 642\\
       7& 40-100  INTEGRAL &  1.00E-12&2009A\&A, 505, 417\\
       9$^2$& 20-80  BeppoSAX & 1.50E-11&2004 ApJ, 602, L73\\
      10$^2$& 20-80  BeppoSAX&  2.30E-11 \\
      11$^2$& 20-80  BeppoSAX&  1.30E-11\\
      13$^3$&20-60 INTEGRAL&  1.52E-12&2008 A\&A, 485, 707\\
      14$^2$& 20-50  INTEGRAL & 1.60E-11&2005 ApJ, 625, 89\\
      15$^1$& 15-55  Swift    & 2.33E-11&2009 ApJ, 690, 367\\
      16$^4$& 10-50       &       1.67E-13&2008 A\&A, 484, 51 \\

      17$^3$& 20-40  INTEGRAL&  1.10E-11 \\
      18$^2$ &20-40  INTEGRAL & 1.44E-11&2006 ApJ, 636, 765\\
      19$^2$& 20-40  INTEGRAL  &1.09E-11&2006 ApJ, 638, 642\\
      21$^2$& 2-10  INTEGRAL &  8.82E-11 &\\
      22$^4$& 2-10  ASCA &      3.20E-12&2004 ApJ, 609, 603\\
      23$^2$ &2-10  BeppoSAX &  9.80E-11 &2007 A\&A, 472, 705\\
      24$^3$ &2-7  XMM-PN  &    3.83E-11&2010 A\&A, 523, 22\\
      25$^3$ &2-7 XMM-MOS1&    4.29E-11 \\
      26$^3$ &2-7 XMM-MOS2 &   4.23E-11\\
      27$^3$& 2-7  Chandra&     4.34E-11\\
      29$^4$& 0.5-2  ROSAT&     1.82E-10 &2004 ApJ, 601, 610\\
      30$^2$& 0.1-2.4  ROSAT&   3.19E-10&2009 ApJ, 690, 879\\
      31$^4$& 0.5-2 ROSAT&     2.20E-10&2011 A\&A, 526, 79\\
\hline
\multicolumn{4}{l}{Notes:~~~Flux in ergs cm$^{-2}s^{-1}$~~~~~Bandpasses in keV}\\
\multicolumn{4}{l}{$^1$From fitting to map~~~~~$^2$Flux integrated from map or image}\\
\multicolumn{4}{l}{$^3$Model~~~~~$^4$Total flux~~~~Flux uncertainties: footnote 2.}\\
\end{tabular}
\end{table}

\section{Large scale structure \label{sec:lsscont}}
By statistically cross-correlating with tracers of cosmic structure, it is possible to indirectly determine which populations most strongly shape the UGRB. This approach leverages the fact that different source classes---such as blazars, star-forming galaxies, or potential dark matter annihilation or decay---trace the large-scale structure in distinct ways. By comparing spatial correlations across energy bands and redshift distributions, one can isolate the contribution of each population. Additionally, cross-correlation mitigates certain instrumental systematics and foreground contaminations, enhancing sensitivity to faint, unresolved sources. Moreover, this method offers a promising avenue to probe dark matter (DM). If DM particles annihilate or decay into Standard Model particles, they are expected to produce $\gamma$-rays that trace the underlying matter distribution. A statistically significant correlation between the $\gamma$-ray sky and the cosmic density field could thus hint at a DM-induced component. By combining cross-correlation measurements with theoretical models of DM distribution and annihilation channels, it becomes possible to place competitive constraints on DM properties such as its mass and annihilation cross-section. Commonly used tracers of cosmic structure include phenomena such as weak gravitational lensing (\cite{camera2013novel,camera2015tomographic,shirasaki2014cross,shirasaki2016cosmological,troster2017cross,DES:2019ucp}), 
galaxy (\cite{xia2015tomography,Cuoco2015,regis2015particle,paopiamsap2024constraints}) and galaxy cluster clustering (\cite{colavincenzo2020searching,branchini2017cross,shirasaki2018correlation,hashimoto2019measurement,tan2020bounds}), and lensing of the Cosmic Microwave Background (CMB) (~\cite{fornengo2015evidence}). These observables probe the large-scale matter distribution across cosmological distances, making them powerful tools for investigating the nature of dark matter and for gaining deeper insights into the astrophysical components of the UGRB.

\begin{figure}
     \centering
     \includegraphics[width=\linewidth]{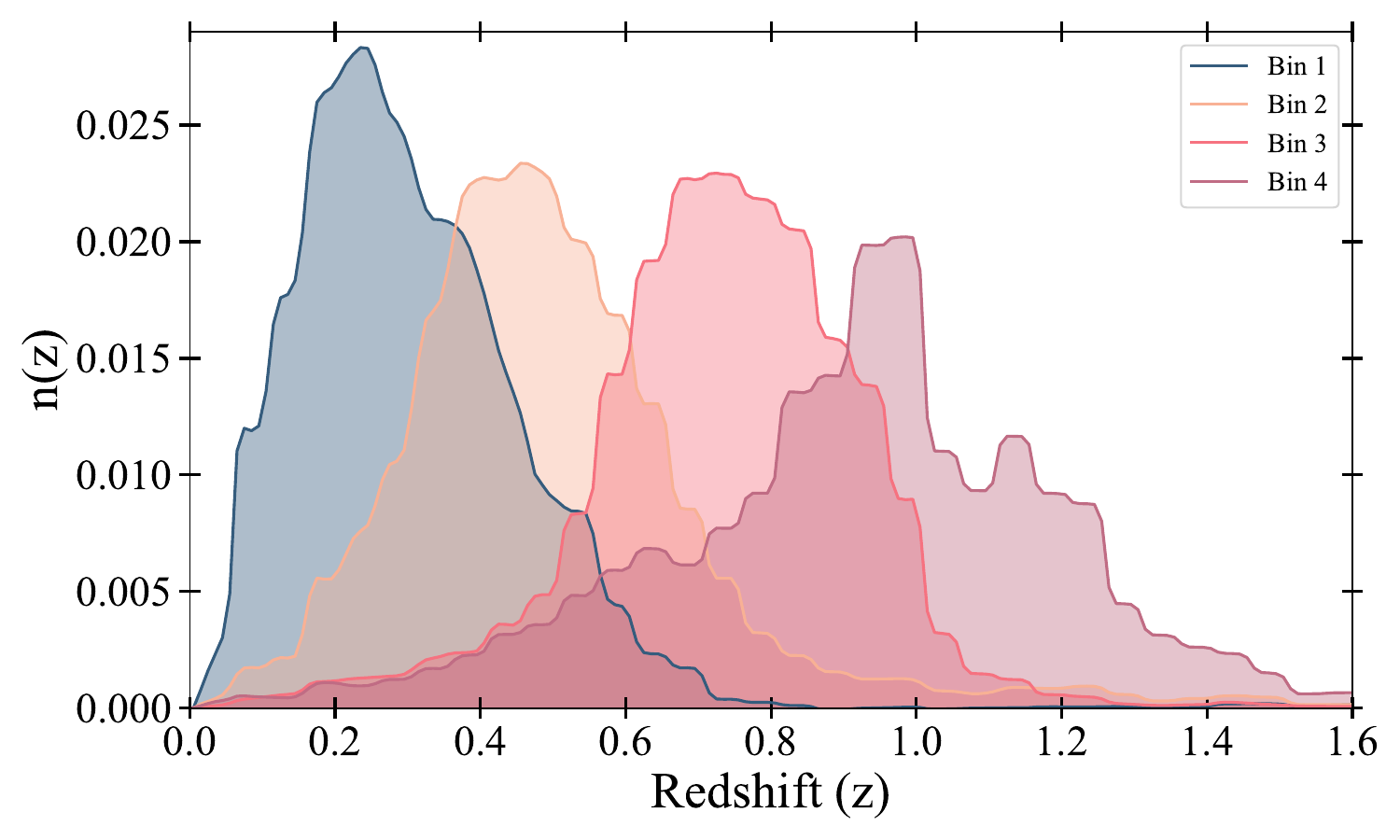}
     \caption{The normalised redshift distribution $n(z)$ of the DES Y3 source galaxies, spanning a weak lensing area of 4143 $\mathrm{deg}^2$, with a total of 100 million galaxies. The data for the redshift distribution is taken from the publicly available DES database: \url{https://des.ncsa.illinois.edu/releases/y3a2/Y3key-catalogs}. The calculation of the redshift bin centres has been detailed in \cite{myles2021dark}, as $\langle z_1 \rangle = 0.339$, $\langle z_2 \rangle = 0.528$, $\langle z_3 \rangle = 0.752$, and $\langle z_4 \rangle = 0.952$. }
     \label{fig:pz}
 \end{figure}

\begin{figure}[!h]
    \centering
    \includegraphics[width=0.50\textwidth]{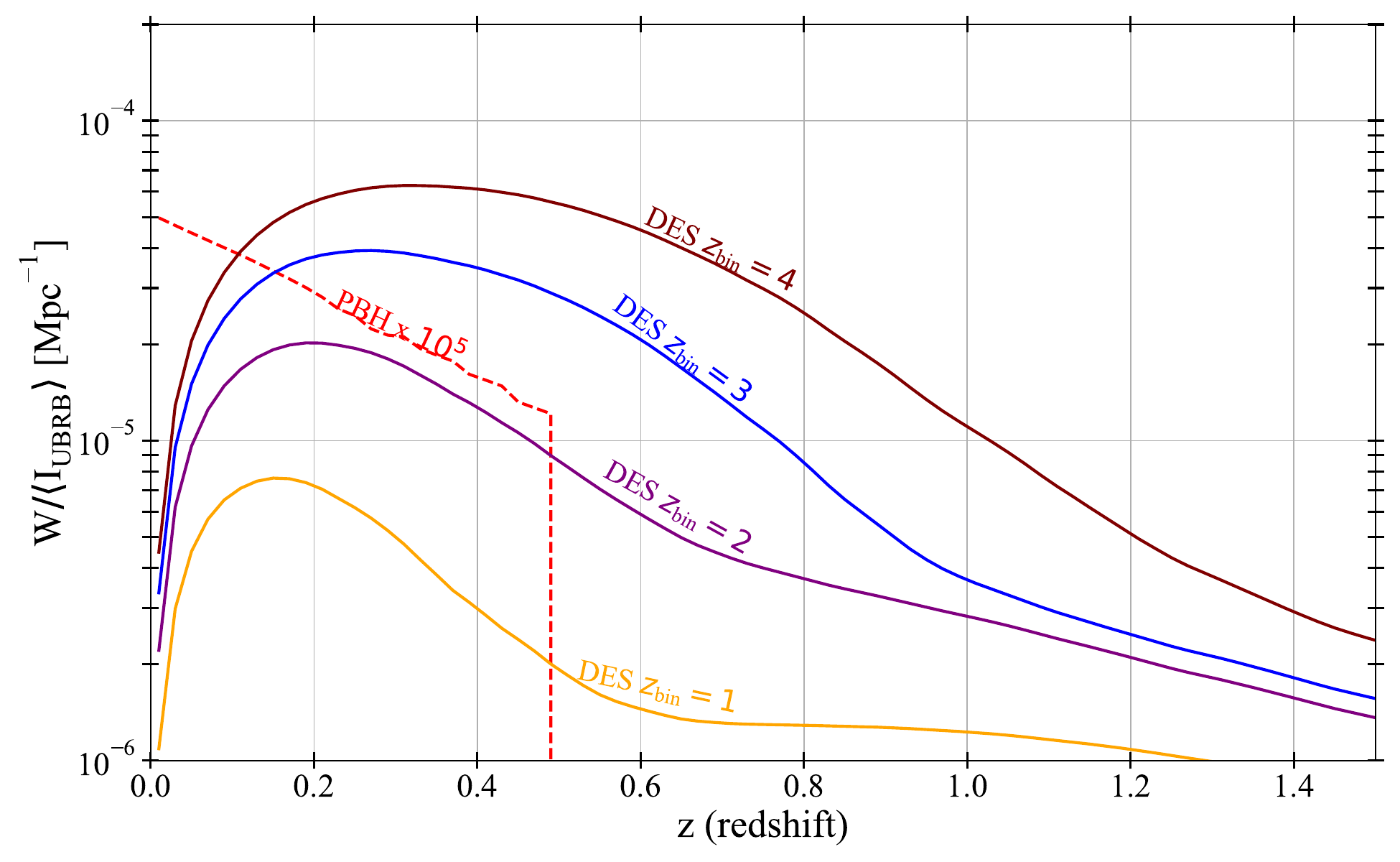}
    \caption{The window function of Primordial Black Hole evaporation in $\gamma$-rays (dashed red) compared to that of the DES Y3 lensing window functions (solid lines) up to redshift $z = 1.5$. The PBH window function is computed in the energy bin 2.290-4.786 GeV and normalized by the average UGRB intensity at the same energy.}
    \label{fig:PBH_Shear_Wfunc}
\end{figure}

 \cite{thakore2025high} have analyzed cross-correlations between the UGRB and the matter distribution in the Universe as traced by gravitational lensing, utilizing 12 years of observations from the Fermi Large Area Telescope (Fermi-LAT) in combination with 3 years of weak lensing data from the Dark Energy Survey (DES) \footnote{A full analysis of the cross-correlation pipeline and its concomitant modelling procedures can be found in \cite{thakore2025high}}, across 9 energy bins between 0.631 to 1000 GeV, 12 angular bins between 5 to 600 arcmins, and a photometric redshift range for the source galaxies as shown in Fig. \ref{fig:pz}.  
 From the aforementioned cross-correlations, they  detect a correlation with a signal-to-noise ratio of 8.9. A majority of the statistical significance of this signal appeared to originate at large scales, thereby  demonstrating that a substantial portion of the UGRB aligns with the mass clustering of the Universe traced according to weak lensing. This could be due, in part, to Hawking radiation from PBHs residing in DM halos. To probe the feasibility of constraining the primordial black hole (PBH) fraction in large-scale structure via cross-correlation techniques, we examine the redshift overlap between the PBH window function and the DES Y3 tangential shear window function, as illustrated in Fig.~\ref{fig:PBH_Shear_Wfunc}. This overlap is critical for determining the sensitivity of weak lensing to potential PBH-induced gamma-ray signals. In our modeling, we consider the photon flux emitted from evaporating PBHs, adopting the mass loss rate derived in \cite{mos}, which accounts for both photon and particle emission channels. The extent of overlap in the window functions directly informs the potential for detecting a cross-correlation signal and, by extension, constraining the PBH abundance.

For the PBH window function, we have adopted a scenario where the PBH dark matter fraction is set to $f = 1$, in order to estimate the maximum possible $\gamma$-ray flux from evaporating PBHs of mass $10^{-18}\ \mathrm{M_\odot}$. This choice provides an absolute upper bound on the expected emission signal and allows us to assess the detectability of such a scenario in the context of cross-correlations. As outlined in Sec.~\ref{sec:srcemission}, the PBH mass distribution is assumed to follow an initial mass function (IMF) of the form $n(m) \propto m^{-1}$, representing a scale-invariant distribution commonly considered in the literature. Cosmological parameters used in the modeling of the window functions are taken from the Planck 2018 results \cite{aghanim2020planck}, ensuring consistency with the assumptions employed in \cite{thakore2025high}.

The redshift-dependent PBH window function $W_{\mathrm{PBH}}$ is given by:

\begin{equation}
    W_{\mathrm{PBH}} = \int_{\Delta E} c^2 f \cdot \frac{dm}{dt} \cdot \frac{dt}{dz} \cdot \frac{e^{-\tau}}{(1+z)^4} \, dz,
\end{equation}

where $dm/dt = \alpha \hbar c^4 / (G m)^2$ denotes the Hawking evaporation rate, $f$ is the fractional contribution of PBHs to the total dark matter density, and $t$ is the cosmic time at redshift $z$. The factor $e^{-\tau}$ accounts for attenuation due to the optical depth $\tau$, integrated over the line of sight for the DES Y3 source galaxy redshift distribution. The $(1+z)^{-4}$ term accounts for redshifting of the energy flux and cosmological dimming.

The corresponding average gamma-ray intensity from PBH evaporation can be expressed as:

\begin{equation}
\langle I_{\mathrm{avg}} \rangle = \int d\chi\, W_{\mathrm{PBH}} = \int_{\Delta E} dE \int dz \frac{c}{H(z)} W_{\mathrm{PBH}}, 
\label{eq:Iavg}
\end{equation}

where $\chi$ is the comoving radial distance, and $H(z)$ is the Hubble parameter at redshift $z$. Evaluating Eq.~\ref{eq:Iavg}, we find that the average PBH intensity at $f = 1$ and $m = 10^{-18}\ \mathrm{M_\odot}$ peaks at $\sim 5$~GeV and yields a flux of $\langle I_{\mathrm{avg}} \rangle \sim 10^{-13}\ \mathrm{cm^{-2}\,s^{-1}\,sr^{-1}}$. In comparison, the measured intensity of the unresolved gamma-ray background (UGRB) is approximately $\langle I_{\mathrm{avg}} \rangle \sim 10^{-8}\ \mathrm{cm^{-2}\,s^{-1}\,sr^{-1}}$. 

This stark contrast spanning about five orders of magnitude indicates that PBHs, even under optimistic assumptions, contribute negligibly to the total UGRB flux. The dominant contributors to the UGRB are known to be blazars and other astrophysical sources (see e.g. \cite{di2018deriving, korsmeier2022flat}), and matching the observed UGRB intensity would require an unphysically high PBH abundance.

Additionally, as shown in Fig.~\ref{fig:PBH_Shear_Wfunc}, the PBH window function exhibits limited overlap with the DES Y3 shear window function. Specifically, the PBH contribution falls off sharply at $z \gtrsim 0.5$, whereas the lensing signal from DES is more prominent at higher redshifts. This lack of overlap in redshift space reduces the efficiency of a cross-correlation approach. These combined findings imply that a cross-correlation between the UGRB and DES Y3 tangential shear is unlikely to yield meaningful constraints on PBH evaporation for this mass range.

\section{Draco X}
\begin{figure}
    \includegraphics[width=.5\textwidth]{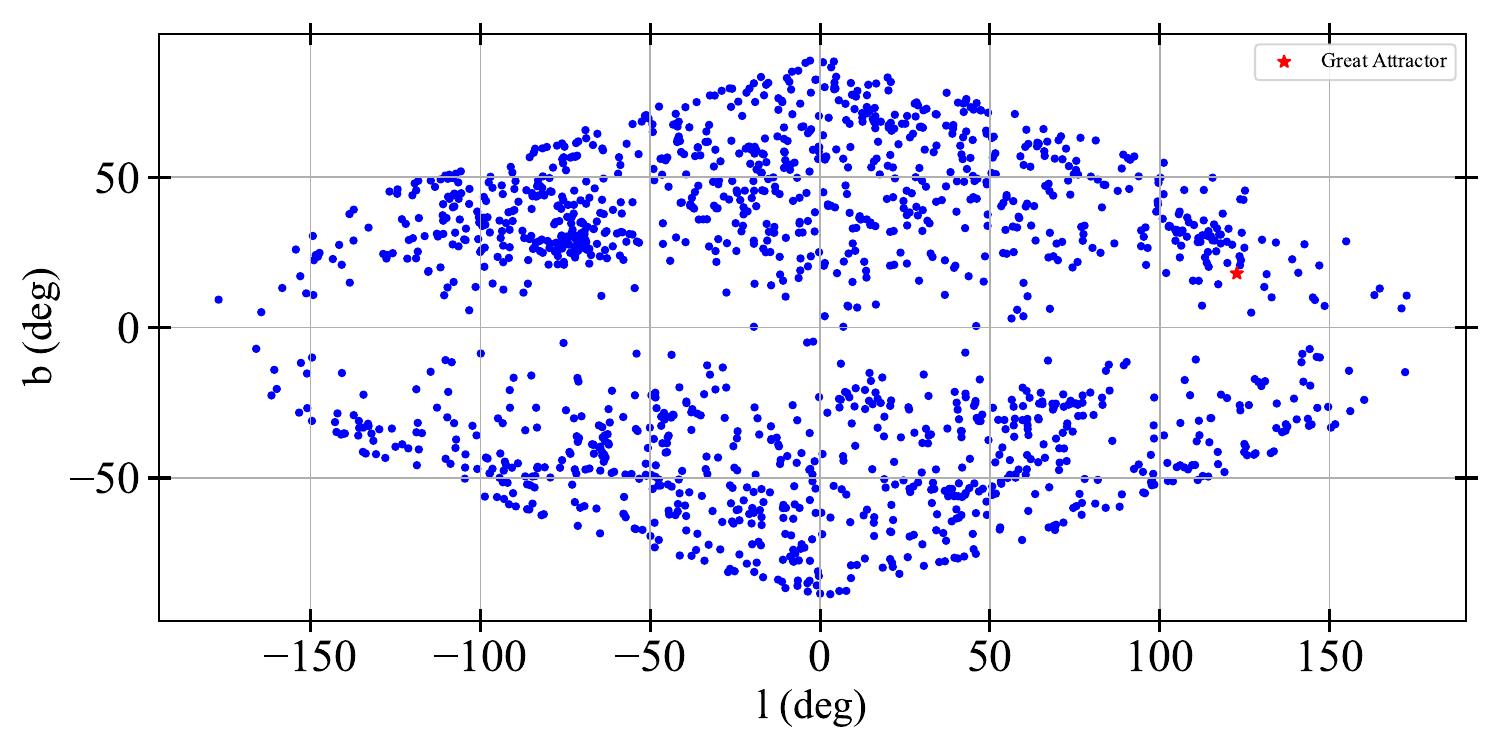}
    \caption {X-ray clusters in the MCXC-II catalog with redshift $<$ 0.2. This cut was made to improve completeness in the southern celestial hemisphere. The long axis is Galactic longitude, the short axis, Galactic latitude. The red dot in the first quadrant is the Great Attractor \citep{dress}, 
    partly obscured by the Galactic plane. The
    nearby concentration of x-ray clusters form part
    of the superstructure Quipi according
    to \cite{boh}.}
    \label{fig:mcxcii}
\end{figure}

\begin{figure}[h]
    \includegraphics[width=.5\textwidth]{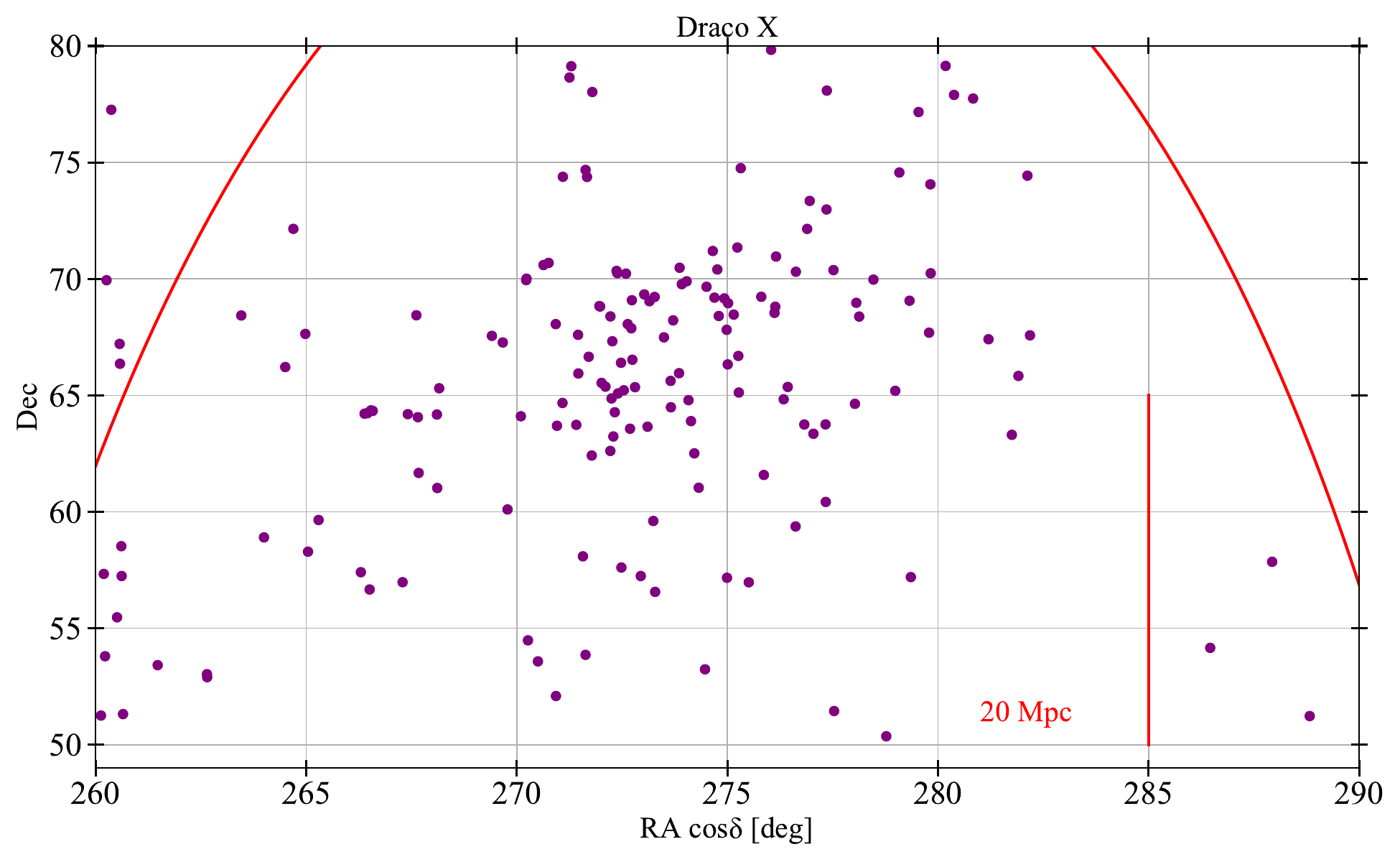}
    \includegraphics[width=.5\textwidth]{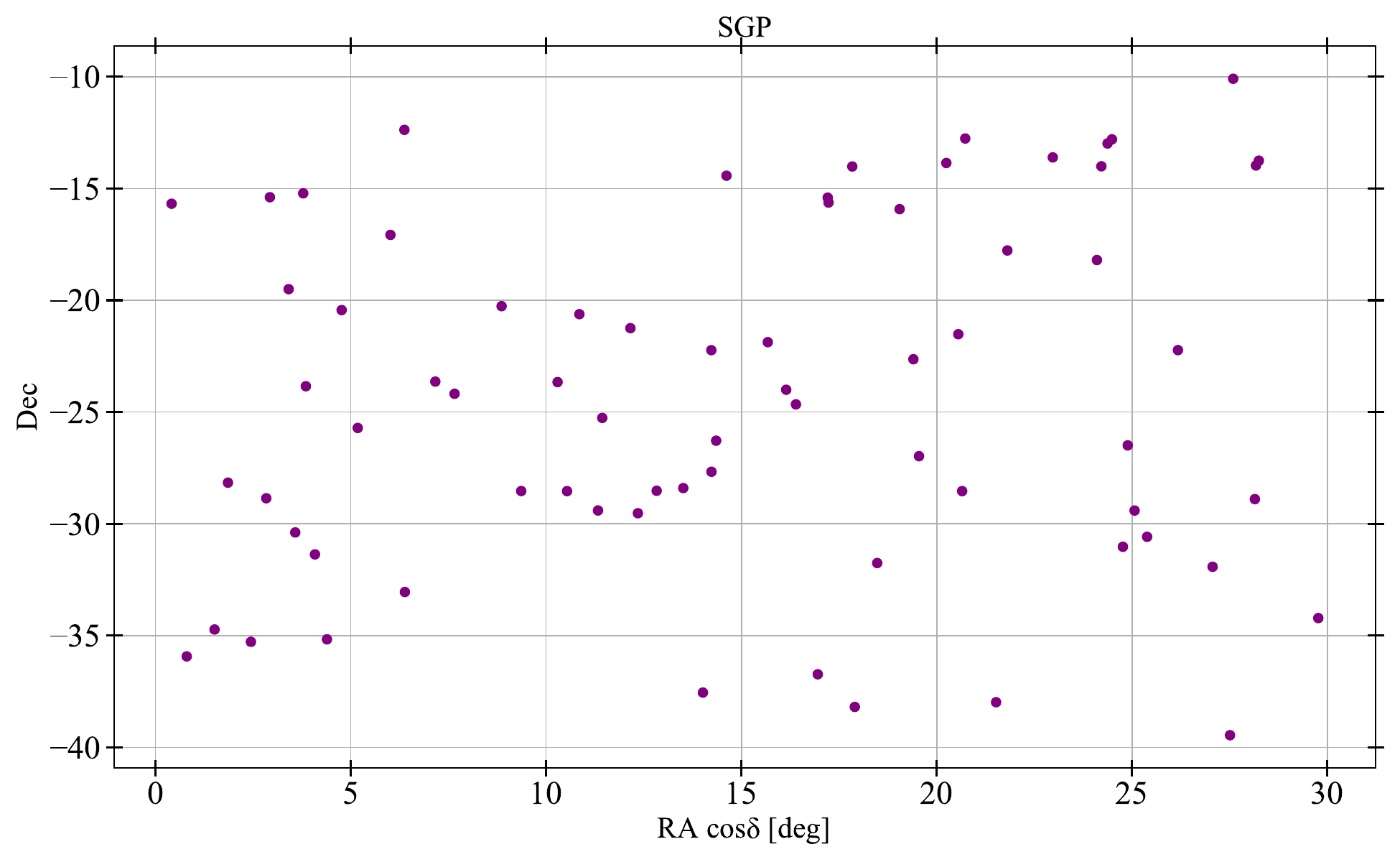}
    \caption {X-ray clusters in the MCXC-II catalog enlarged from the previous figure. A control sample is shown below it around the south galactic pole. The red
    arcs are lines of constant RA.}
    \label{fig:dracox}
\end{figure}

\cite{sad} 
have compiled a catalog of x-ray clusters whose distribution is shown in Figure \ref{fig:mcxcii}.
A prominent cluster of clusters stands out in the second quadrant, which we refer to as Draco X. Its Right Ascension (RA) is 277.3$^\circ$, Declination (Dec) 73.2$^\circ$, and its observed surface overdensity, $\delta\rho/\rho$, is approximately 5, indicating a significant local enhancement in the X-ray cluster density compared to the cosmic average. A lower limit on its X-ray luminosity, relative to the well-studied Coma Cluster, is estimated to be 58 times greater, suggesting a remarkably energetic system. Figure \ref{fig:dracox} is an enlargement. The integrated mass within the inner 100 square degrees of Draco X is approximately 7.2 $\times$ 10$^{16}$ M$_\odot$ (with a 90\% confidence interval ranging from 0.66 10$^{16}$ to 2.0 $\times$ 10$^{16}$ M$_\odot$). This makes its mass roughly a tenth of the estimated mass of the Shapley Supercluster, as determined by \cite{boh}. The physical extent of Draco X is a significant fraction of the cosmic microwave background (CMB) sound horizon at its redshift, a characteristic not entirely unexpected among the declining peaks of the CMB power spectrum, which reflect the largest structures that could form in the early universe.

The redshift of Draco X is z = 0.12, with a redshift dispersion of 0.056. If the Coma Cluster serves as a representative model for galaxy clusters, the expected $\gamma$-ray flux from Draco X would be over twice that observed from Coma, due to its higher luminosity and potentially more active internal processes. However, a comparison with the observed overdensity of extended sources (with semimajor axes greater than 0.25 degrees) in the Fermi-LAT DR3 4FGL catalog \citep{abd}. reveals a background overdensity of $\delta\rho/\rho$ = 0, indicating no significant enhancement of extended $\gamma$-ray sources in the direction of Draco X. Furthermore, our analysis of the coincidence rate between these Fermi-LAT sources and Draco X clusters shows a rate two to three times higher than expected by chance alone. Evidently, the statistical correlation between X-ray clusters and extended Fermi-LAT sources, at least for this sample, appears to be weak or absent, suggesting distinct emission mechanisms or source populations.

With an estimated cluster-cluster interaction rate of 1 per 2 Gyr, it is highly probable that some of the gaseous spheres and gravitational potential wells of the constituent clusters within Draco X have undergone merger events, similar to the famous Bullet Cluster (1E 0657-56). The observed spatial separation between the X-ray emitting gas (which is collisional) and the dark matter (which is largely collisionless) in such events provides crucial constraints on the ratio of the self-interaction cross-section to the mass for self-interacting dark matter (\citep{Ran}). Given its immense mass, high density, and evident ongoing dynamics, Draco X would be an exceptionally interesting and valuable environment for detailed studies of such collision-induced phenomena, offering unique insights into the fundamental properties of dark matter.

\par
\par\vspace{1cm}
\par

\section{Conclusions}
Our results underscore both the opportunities and limitations in using high-energy observations to probe fundamental physics. The x-ray–bright Coma cluster is a 
candidate for detection of Hawking radiation GeV flux, but, until PBH are detected unequivocally,  such an interpretation is speculative and limited by uncertainties in the modelling of intracluster emission. More distant x-ray structures, such as the newly identified Draco X, show no corresponding GeV emission. 
Like the closer Great Attractor, its gravitational attraction would be enough to create peculiar velocities of $\gtrsim$500 \kms at 100 Mpc distance.

At higher redshifts, we also attempted to constrain PBH evaporation through cross-correlations between the UGRB and tangential shear, following the framework outlined by \citet{thakore2025high}. While the method remains effective for investigating astrophysical sources such as blazars, we found that the lack of substantial overlap between the PBH and shear window functions under current assumptions weakens its sensitivity to PBH contributions. Extending these analyses with forthcoming data from DESI, as suggested by \citet{Zhou}, may help to improve statistical power and enable broader tests of both astrophysical and exotic components of the UGRB.

.

To briefly summarize,

\begin{itemize}
    \item Given its x-ray emission and dark matter, Hawking radiation is a possible contributor to the Coma cluster's GeV flux. 
    \item We see no correlation between PBH emission that might be inferred from the DES DM distribution and the UGRB.
    \item Among more distant x-ray clusters we discovered a cluster of clusters, Draco X, but it is not seen at GeV energies. 
Like the Bullet cluster,
Draco X may serve as a useful site for studying cluster mergers and DM behaviour. 
\end{itemize}


\bibliography{bibliography}%
\bibliographystyle{mnras}
\section*{Acknowledgement}
The ARC Centre of Excellence for Dark Matter Particle Physics is funded by the Australian Research Council with grant CE200100008. 
This research has made use of the NASA/IPAC Extragalactic Database (NED),
which is operated by the Jet Propulsion Laboratory, California Institute of Technology,
under contract with the National Aeronautics and Space Administration.

BT acknowledges the support from the  Research grant TAsP (Theoretical Astroparticle Physics) funded by \textsc{infn}. We acknowledge the assistance provided by Marco Regis in the analysis part of Sec. \ref{sec:lsscont}.
\end{document}